\documentclass[aps,prd,amssymb,nofootinbib,twocolumn]{revtex4-1}
\usepackage{amsmath}
\usepackage{amssymb}
\usepackage{graphicx}
\usepackage{array}
\usepackage[usenames]{xcolor} 
\usepackage{natbib}
\usepackage[caption=false]{subfig}
\usepackage{xstring}
\usepackage{refcount}
\usepackage{ulem}

\newcommand{\be}{\begin{equation}}
\newcommand{\ee}{\end{equation}}
\newcommand{\eqcite}[1]{Eq.~(\ref{#1})}

\newcommand{\figcite}[1]{\IfInteger{\getrefnumber{#1}}{Fig.~\ref{#1}}{Fig.~\StrGobbleRight{\getrefnumber{#1}}{1}(\StrRight{\getrefnumber{#1}}{1})}}

\newcommand{\fignumcite}[1]{\IfInteger{\getrefnumber{#1}}{\ref{#1}}{\StrGobbleRight{\getrefnumber{#1}}{1}(\StrRight{\getrefnumber{#1}}{1})}}

\begin{document}

	\title{Upper limit set by causality on the tidal deformability of a neutron star}
	\author{Eric D. \surname{Van Oeveren}}
	\author{John L. Friedman}
	\affiliation{Leonard E. Parker Center for Gravitation, Cosmology and Astrophysics \\ 
	University of Wisconsin-Milwaukee \\
	3135 N Maryland Ave, Milwaukee, WI 53211, USA}

	\begin{abstract}
	A principal goal of gravitational-wave astronomy is to constrain the neutron star equation of state (EOS) by measuring the tidal deformability of neutron stars.  The tidally induced departure of the waveform from that of point-particle (or spinless binary black hole (BBH)) increases with the stiffness of the EOS.  We show that causality (the requirement that the speed of sound is less than the speed of light for a perfect fluid satisfying a one-parameter equation of state) places an upper bound on tidal deformability as a function of mass.  Like the upper mass limit, the limit on deformability is obtained by using an EOS with $v_{\rm sound} = c$ for high densities and matching to a low density (candidate) EOS at a matching density of order nuclear saturation density.  We use these results and those of [B.D. Lackey \textit{et al.}, Phys. Rev. D \textbf{89}, 043009 (2014)] to estimate the resulting upper limit on the gravitational-wave phase shift of a black hole-neutron star (BHNS) binary relative to a BBH. Even for assumptions weak enough to allow a maximum mass of $4 M_\odot$ (a match at nuclear saturation density to an unusually stiff low-density candidate EOS), the upper limit on dimensionless tidal deformability is stringent. It leads to a still more stringent estimated upper limit on the maximum tidally induced phase shift prior to merger.   We comment in an appendix on the relation between causality, the condition $v_{\rm sound} < c$, and the condition $dp/d\epsilon < 1$ for the effective EOS governing the equilibrium star.  

	\end{abstract}

\maketitle
\section{Introduction}
As the second generation of ground-based interferometric graviational wave detectors 
(Advanced LIGO, Advanced Virgo, KAGRA, and LIGO-India) approach design sensitivity, 
we are likely to detect each year the inspiral and coalescence of several compact binary 
systems that include neutron stars, both BHNS and binary neutron-star (BNS) systems.  
These observations can constrain the neutron-star equation of state (EOS), which gives the pressure $p$ in terms of the energy density $\epsilon$. A stiffer EOS, where the pressure increases rapidly with density, yields stars with larger radii and larger tidal effects on the waveform, 
governed by the star's {\it tidal deformablity}.    
In particular, tidal distortion during inspiral increases with the stiffness of 
the EOSs. Because energy is lost both to gravitational waves and to the work needed 
to distort the stars, the inspiral proceeds more rapidly for stars with greater 
tidal deformability.  The result is a waveform in which the increase in frequency is more 
rapid and in which coalescence occurs sooner -- at lower frequency. 

Beginning with work by Kochanek \cite{Kochanek} and Lai and Wiseman \cite{LW}, a number of 
authors have studied the effect of tides on inspiral waveforms. Simulations 
\cite{PBKLS16,CYKCY15,ZCM,USE,FGRT,FGR,BGGHTZ,GBBGHLTZ,STU,SU,OJ,TG,S,SU2,ST,STU2,MGS,MDSB,LSET} of BHNS 
and BNS systems  and analytic approximations in the context of post-Newtonian theory \cite{VFH} and the 
Effective-One-Body (EOB) formalism \cite{HTF,BDN15,BNDD15} are nearing the precision needed to extract 
neutron-star deformability from observations with the projected sensitivity of Advanced LIGO.  Recent estimates of the measurability of tidal effects and the ability of these observatories to constrain the EOS with signals from BHNS and BNS systems 
are given in \cite{HKSS16,DNV,DLAVV,AMDLTVVV15,WCOLFLR} and references therein.

In this work, we obtain the upper limit imposed by causality on the tidal deformability of neutron 
stars and estimate the resulting constraint on the maximum departure of the waveform of a BHNS 
inspiral from a corresponding spinless BBH inspiral.\footnote{After this paper was posted to arXiv, Moustakidis \cite{Moustakidis} pointed out a preprint by him and his coauthors that also obtains upper limits on neutron star mass and tidal deformability imposed by bounds on the speed of sound, including $v_{\mathrm{sound}} \leq c$. However, they use a matching density (described in the next section) 50\% higher than ours, giving less conservative results. Furthermore, they do not consider tidal effects during late inspiral, whereas we apply the results of \cite{Lackey} to do so.} The limit is 
analogous to the upper limits on neutron-star mass $M_{NS}$ \cite{RR,Obese} and radius $R$ \cite{Lattimer}. In each case, one assumes an EOS of the form $p = p(\epsilon)$ that is known below an energy density $\epsilon_{\mathrm{match}}$, and one obtains a limit on $M$ and $R$ by requiring that the EOS be causal for $\epsilon>\epsilon_{\mathrm{match}}$ in the sense that the sound speed, given by $\sqrt{dp/d\epsilon}$, must be less than the speed of light.  Because the sound speed is a  
measure of the stiffness of the EOS, this is a constraint on the stiffness. 
An upper limit on tidal deformability then implies an upper limit on the departure of 
gravitational wave phase shifts from corresponding waveforms of BBH inspiral.   

We use metric signature $-+++$ and gravitational units with $G=c=1$.

\section{Method}
\subsection{Causal EOS}
 
For a perfect fluid with a one-parameter EOS $p=p(\epsilon)$, causality implies 
that the speed of sound, $\sqrt{dp/d\epsilon}$, is less than the speed of light.
That is, the dynamical equations describing the evolution of fluid and metric are hyperbolic, with 
characteristics associated with fluid degrees of freedom lying outside the light cone unless 
\begin{equation}
	\frac{dp}{d\epsilon} \leq 1.
\label{e:ineq0}\end{equation}
There is some inaccuracy in using the one-parameter EOS that governs the equilibrium 
star to define the characteristic velocities of the fluid, because fluid oscillations 
with the highest velocities have frequencies too high for the temperature of a fluid 
element and the relative density $Y_i$ of each species of particle to reach their values for the 
background fluid star.  Nevertheless, using a result of 
Geroch and Lindblom \cite{GL91}, we show in Appendix A that causality implies the 
equilibrium inequality (\ref{e:ineq0}) for locally stable relativistic fluids 
satisfying a two-parameter EOS $p = p(\epsilon,s)$, 
where $s$ is the entropy per baryon.  For the multi-parameter equation of state 
$p = p(\epsilon, s, Y_i)$, with $Y_i$ the relative density of each species of 
particle, one must assume without proof that causality implies $v_{\rm sound}<1$; 
the equilibrium inequality (\ref{e:ineq0}) again follows from local stability.  
   
The speed of sound is a measure of the stiffness of the EOS.  The well-known upper 
limit on the mass of a neutron star and a corresponding upper limit on its radius are obtained 
by using the stiffest EOS consistent with causality and with an assumed known form 
at low density.  That is, above a density $\epsilon_{\rm match}$, the EOS is given by 
\be
   p - p_{\mathrm{match}} = \epsilon - \epsilon_{\mathrm{match}}, 
\label{e:causal_eos}\ee 
where $p_{\rm match}$ is fixed by continuity to be the value of $p$ at $\epsilon_{\rm match}$ 
for the assumed low-density 
EOS.  The upper limits on mass and radius are then found as functions of 
the matching density $\epsilon_{\rm match}$.  

In this work, we again use an EOS 
of this form to find an upper limit on neutron-star deformability. 
To be conservative, as our low-density EOS we choose  
the MS1 EOS \cite{MS1}, which is among the stiffest candidate equations of 
state.  Our {\it matched causal EOS} is then given by  
\begin{equation}
	p(\epsilon) = \begin{cases} p_{\mathrm{MS1}}(\epsilon), & \epsilon \leq \epsilon_{\rm match} \\
	\epsilon - \epsilon_{\rm match} + p_{\mathrm{MS1}}(\epsilon_{\rm match}), & \epsilon \geq \epsilon_{\rm match}. \end{cases}
\label{e:EOS}
\end{equation}
In computing the deformability, we consider only irrotational neutron stars; and in 
estimating the effect of tides on the inspiral phase, we neglect 
resonant coupling of tides to neutron-star modes.  Tidal deformation of slowly rotating 
relativistic stars is treated by Pani {\it et al.} \cite{PGMF15}; and  
Essick {\it et al.} \cite{EVW16} argue that tidal excitation of coupled modes may alter the waveform in BNS systems.    

\subsection{Static, Spherical Stars}
We next construct the sequence of static spherical stars based on the causal EOS (\ref{e:EOS}).
We numerically integrate the Tolman-Oppenheimer-Volkoff (TOV) equation \cite{OV},
\begin{equation}
\left(1-\frac{2m}{r}\right) \frac{dp}{dr} = -\frac{1}{r^2}(\epsilon + p)(m+4\pi r^3 p),
\label{TOV}
\end{equation}
where $m(r)$ is the total mass-energy inside radius $r$, related to $\epsilon$ by
\begin{equation}
\frac{dm}{dr} = 4\pi r^2 \epsilon.
\label{mprime}
\end{equation}
A member of the sequence is specified by its central density $\epsilon_c$.  Its circumferential 
radius $R$ is the value of the Schwarzschild coordinate $r$ at which $p(r)=0$, and its gravitational mass 
is $M=m(R)$. 

	\subsection{Calculating the Tidal Deformability}
      
     The departure of the inspiral of a BHNS binary from point-particle (or spinless BBH) inspiral depends on the tidal deformation of the neutron star due to the 
tidal field of its companion.  For large binary separation, the metric near the 
neutron star can be written as a linear perturbation of the Schwarzschild metric 
of the unperturbed star that has two parts: The tidal field of the 
companion, expressed in Schwarzschild coordinates about the center of mass of the 
neutron star, has the form of an external quadrupole field; and the induced 
quadrupole distortion of the neutron star gives a second quadrupole contribution to 
the perturbed metric.  That is, outside the support of the star, the 
quadrupole perturbation is a sum, 
\be
  \delta g_{\alpha\beta} = \delta_{\rm external}\, g_{\alpha\beta} + \delta_{\rm induced}\, g_{\alpha\beta},
\ee
of two time-independent solutions to the field equations linearized about a 
vacuum Schwarzschild geometry. In a gauge associated with asymptotically Cartesian and 
mass centered coordinates, the contributions to the perturbed metric have the form 
\begin{align}
  \delta_{\rm external}\, g_{tt} &= -r^2 \mathcal{E}_{ij} n^i n^j + \mathcal{O}(r),  
\end{align}
with no $r^{-3}$ contribution, and 
\begin{align}
  \delta_{\rm induced}\, g_{tt} &=  \frac{3}{r^3} Q_{ij}\left(n^i n^j - \frac{1}{3} \delta^{ij}\right) + \mathcal{O}(r^{-4}).    
\end{align}
Here $n^i = x^i/r$ is an outward-pointing unit vector, $\mathcal{E}_{ij}$ is the tracefree tidal field from the black hole, and $Q_{ij}$ is the neutron star's induced quadrupole moment.  The quadrupole moment tensor
$Q_{ij}$ is proportional to $ \mathcal{E}_{ij}$, 
	\begin{equation}
	Q_{ij} = -\lambda \mathcal{E}_{ij},
	\label{td}
	\end{equation}
and the constant of proportionality $\lambda$ is 
the {\it tidal deformability} of the neutron star. It measures the  
magnitude of the quadrupole moment induced by an external tidal field
and is proportional to the 
(dimensionless) $\ell=2$ tidal Love number \cite{FH}
	\begin{equation}
	k_2 = \frac{3\lambda}{2R^5}.
	\label{k2_def}
	\end{equation}

	After constructing the one-parameter family of spherical stars satisfying Eqs.~(\ref{e:EOS}), (\ref{TOV}), and (\ref{mprime}),  we tidally perturb them, compute $k_2$ and the radius $R$ of each star, and then find the tidal deformability $\lambda$ from Eq.~(\ref{k2_def}).   To calculate $k_2$, we use the method described by Hinderer \cite{Hinderer}: A perturbation of the spherically symmetric background metric
 \be
       \mathbf{g} = -e^{2\nu}dt^2+  \frac{1}{1-2m/r}dr^2+ r^2(d\theta^2+ \sin^2\theta d\phi^2),
 \ee
	with $\nu(r)$ determined by
	\begin{equation*}
	\left(1-\frac{2m}{r}\right)\frac{d\nu}{dr} = \frac{1}{r^2}(m + 4\pi r^3 p),
	\end{equation*}
	is found in the Regge-Wheeler gauge \cite{RW}, with $\delta \mathbf{g}$ a linear, quadrupolar, static, polar-parity perturbation given by \cite{Hinderer} \footnote{Note that, because this gauge does not conform to the constraints of an asymptotically Cartesian and mass-centered chart, there are additional terms in the expansion of the asymptotic metric.} 
\begin{align}
    \delta \mathbf g 
	   =& (-e^{2\nu} dt^2 +\frac{1}{1-2m/r} dr^2) H\, Y_{2,m}(\theta,\phi)
		\nonumber\\
            &+ r^2 (d\theta^2+\sin^2\theta d\phi^2) K\, Y_{2,m}(\theta,\phi),
\end{align}
where $H$ and $K$ are both functions of $r$. The perturbed Einstein equation gives a differential equation for $H$ \cite{Hinderer}:
	\begin{align*}
	0=&\frac{d^2H}{dr^2}\left(1-\frac{2m}{r}\right) + \frac{dH}{dr}\left[ \frac{2}{r} -  \frac{2m}{r^2} + 4\pi r (p-\epsilon)  \right] \\
	  & -\! H \left[\frac{6}{r^2}-4\pi \left(5\epsilon +9p + \frac{\epsilon + p}{dp/d\epsilon} \right) \right.\\
	  & \qquad \ \ \left. +  4\left(1-\frac{2m}{r}\right)\! \left(\frac{d\nu}{dr}\right)^2\! \right].
	\end{align*}
	In vacuum, $H$ can be written as a linear combination of $P_2^2(r/M-1)$ and $Q_2^2(r/M-1)$, 
where $P_2^2$ and $Q_2^2$ are the $\ell=m=2$ associated Legendre functions. 
When expanded in powers of $M/r$ at infinity, 
$P_2^2(r/M-1) = \mathcal{O}(M/r)^3$ and $Q_2^2(r/M-1) = \mathcal{O}(r/M)^2$. 
The coefficient of $P_2^2$ is therefore related to the quadrupole moment of the star, 
and the coefficient of $Q_2^2$ is related to the tidal field applied by the black hole. 
By matching $H(r)$ and its derivative across the surface of the star, 
one can show  \cite{Hinderer}
	\begin{align}
	k_2 =& \frac{8}{5}C^5 (1-2C)^2 [2+2C(Y-1) -Y]  \nonumber \\ 
	& \times \left\{2C[6-3Y + 3C(5Y-8) + 2C^2(13-11Y) \right.\nonumber \\
	&\hspace{12mm} \left.+ 2C^3(3Y-2) + 4C^4(Y+1)] \right.\nonumber \\
	&\left. + 3(1-2C)^2[2-Y+2C(Y-1)]\log(1-2C) \right\}^{-1},
	\label{k2}
	\end{align}
	where $C=M_{\mathrm{NS}}/R$ is the compactness of the star, \mbox{$Y=R H'(R)/H(R)$}, and $R$ is the radius of the star. Since $k_2$ depends on $Y$ and not $H$ or $H'$ individually, Postinkov, Prakash, and Lattimer \cite{PPL} and Lindblom and Indik \cite{LI2} define
	\begin{equation*}
	y(r) = r \frac{H'(r)}{H(r)},
	\end{equation*}
	which gives rise to the first-order differential equation
	\begin{align}
	\frac{dy}{dr} =& -\frac{y^2}{r} - \frac{r + 4\pi r^3(p - \epsilon)}{r(r-2m)}y
			+ \frac{4(m + 4\pi r^3 p)^2}{r(r-2m)^2}  \nonumber \\
	& + \frac{6}{r-2m} - \frac{4\pi r^2}{r-2m} \left[5\epsilon + 9p + \frac{(\epsilon + p)^2}{\epsilon dp/d\epsilon} \right].
	\label{y}
	\end{align}
	To find $Y = y(R)$, we numerically integrate Eq.~(\ref{y}) and evaluate $y$ at the surface of the star.
	
	Despite appearances, the expression in curly braces in Eq.~(\ref{k2}) is $\mathcal{O}(C^5)$ due to cancellations of terms in curly brackets that are polynomial in $C$ with terms from the expansion of $\log(1-2C)$. For stars of small compactness, calculating $k_2$ directly from Eq.~(\ref{k2}) is difficult because it requires that both the numerator and denominator of the right side are accurately calculated to a large number of decimal places. As a result, we expand $k_2$ to 20 orders in $C$. Since $k_2$ is $\mathcal{O}(C^0)$, this allows for much more accurate results for small $C$. The compactness has a maximum value of $1/2$, so this expansion converges for all stars.
	
	\subsection{Estimating the Gravitational Wave Phase Shift due to Tidal Deformability}
The tidal deformability $\lambda$ defined in the last section accurately describes the actual 
deformation of a neutron star in a binary system only when the neutron star is far from the other compact object. This is for several reasons: As the neutron star approaches the other object, linear perturbation theory and the assumption of a static spacetime used to define $\lambda$ break down; higher-order multipoles in the metric become important; as the star spirals in, its orbital angular velocity increases and becomes comparable to the frequencies of the star's normal modes, and this enhances the star's response to the tidal perturbation \cite{HTF}; and, finally, the neutron star may be tidally disrupted before merger. Nevertheless, the tidal deformability turns out essentially to determine the departure of gravitational waveforms from those spinless BBH 
inspiral in numerical simulations \cite{Lackey,Lackey12,Read13}. 

	A post-Newtonian expansion \cite{VFH} describes the effect of tidal deformability on the phase of the gravitational waveform to linear order in $\lambda$:
	\begin{equation}
	\Delta\Phi_{\mathrm{PN}} = -\frac{3\Lambda}{128\eta}(\pi M f)^{5/3} \left[a_0 + a_1 (\pi M f)^{2/3} \right],
	\label{DeltaPhiPN}
	\end{equation}
	where $\Delta\Phi$ is the difference in gravitational wave phase between a spinless BBH and a BHNS binary, $\Lambda = \lambda/M_{\mathrm{NS}}^5$ is the dimensionless tidal deformability, $M = M_{\mathrm{BH}} + M_{\mathrm{NS}}$ is the total mass of the binary system, $\eta = M_{\mathrm{BH}} M_{\mathrm{NS}}/M^2$ is the symmetric mass ratio, $f$ is the linear frequency of the gravitational radiation, and $a_0$ and $a_1$ are functions of $\eta$:
	\begin{align*}
	a_0 &= 12 [1 + 7\eta - 31\eta^2 - \sqrt{1-4\eta} (1 + 9\eta - 11\eta^2)],\\
	a_1 &= \frac{585}{28} \left[1+ \frac{3775}{234}\eta - \frac{389}{6}\eta^2 
		+ \frac{1376}{117}\eta^3 \right.
	\\&\qquad\quad\left. - \sqrt{1-4\eta} \left(1+ \frac{4243}{234}\eta - \frac{6217}{234} \eta^2 - \frac{10}{9} \eta^3 \right) \right].
	\end{align*}
    Where Eq.~(\ref{DeltaPhiPN}) is valid, in the early inspiral when the frequency $f$ is low, it allows us to easily compute the phase change (the amplitude of the waveform is also affected by tidal deformability, but in this regime the difference in amplitude is small). However, tidal effects are largest during late inspiral when the frequency is high.
	
	To extend the analytic computation to late inspiral, Lackey \textit{et al.} \cite{Lackey} fit the 
amplitude and phase of the gravitational waveforms of neutron star-black hole inspirals 
to the results of numerical simulations, for black hole spins $\chi_{\mathrm{BH}}$ between -.5 and .75, 
and mass ratio $M_{\mathrm{BH}}/M_{\mathrm{NS}}$ in the range 2 to 5. The resulting expressions (below) depend on 
post-Newtonian theory for low frequencies, when the neutron star is still far from the black hole. 
At high frequencies, the fits to numerical results take over:
	\begin{equation}
	A = \begin{cases}
	A_{\mathrm{PN}}, & Mf \leq .01 \\
	A_{\mathrm{PN}}e^{-\eta\Lambda B(\Lambda,\eta,\chi_{\mathrm{BH}}) (Mf - .01)^3}, & Mf > .01
	\end{cases}
	\end{equation}	
	\begin{equation}
	\Delta\Phi = \begin{cases}
	\Delta\Phi_{\mathrm{PN}}(Mf), & Mf \leq .02 \\
	-\eta\Lambda E(\eta, \chi_{\mathrm{BH}}) (Mf - .02)^{5/3} \\
	\quad + \Delta\Phi_{\mathrm{PN}}(.02) \\
	\quad + (Mf-.02)\Delta\Phi_{\mathrm{PN}}'(.02), & Mf > .02.
	\end{cases}
	\end{equation}
	Here the subscript $PN$ indicates the corresponding result from post-Newtonian theory; $B$ is a function of $\Lambda$, $\eta$, and $\chi_{\mathrm{BH}}$; and $E$ is a function of $\eta$ and $\chi_{\mathrm{BH}}$. The parameters of $B$ and $E$ were determined by the numerical fit. In particular,
	\begin{equation*}
	B = e^{b_0 + b_1 \eta + b_2 \chi_{\mathrm{BH}}} + \Lambda e^{c_0 + c_1 \eta + c_2 \chi_{\mathrm{BH}}},
	\end{equation*}
	with 
 $\{b_0,\ b_1,\ b_2\} = \{-64.985,\ \!-2521.8,\ \! 555.17\}$ and $\{c_0,\ c_1,\ c_2\} = \{-8.8093,\ \! 30.533,\ \! 0.6496\}$ as the fitting parameters. Similarly,
	\begin{equation*}
	E = e^{g_0 + g_1 \eta + g_2 \chi_{\mathrm{BH}} + g_3 \eta \chi_{\mathrm{BH}}},
	\end{equation*}
	with \\
\mbox{$\{g_0,\ g_1,\ g_2,\ g_3\}$} = $\{-1.9051,\ \! 15.564,\ \! -0.41109,\ \! 5.7044\}$.
	While high tidal deformabilities increase $|\Delta\Phi|$ relative to a point-particle waveform at a given frequency $f$, they also cause stars to be tidally disrupted earlier in the inspiral, damping the resulting gravitational waves. We define the cutoff frequency $f_{\rm cutoff}$ to be the frequency at which effects from tidal deformation dampen the amplitude by a factor of $e$ relative to the post-Newtonian waveforms. To estimate the total effect of tidal deformability on the phase of the waveform throughout the inspiral, we chose to evaluate $\Delta\Phi$ at $f_{\rm cutoff}$.
	
     The errors in the fitting parameters reported in \cite{Lackey} correspond to errors 
in $\Delta\Phi(Mf_{\mathrm{cutoff}})$ of $\sim\!\! 15\%$ for typical binary parameters. 
The $\Delta\Phi$-values reported below should therefore not be taken as accurate predictions of the 
tidally-induced phase shift. Still, we expect that applying this fit to the matched causal EOS yields 
an upper limit on $|\Delta\Phi|$ with roughly the same error, especially considering the emphasis 
in \cite{Lackey} on avoiding over-fitting and the lower errors reported for larger $\Lambda$-values. 
A more accurate calculation of the phase shift from BHNS or BNS systems with our causal EOS requires 
numerical simulations (now in progress for BNS systems \cite{markakis}) or use of the EOB formalism.
	
	\section{Results}

  Most neutron stars observed by gravitational waves in binary inspiral are likely to have masses in or near 
the $1.25\ M_\odot$ to $1.45\ M_\odot$ range seen in binary neutron star systems, a range 
consistent with formation from an initial binary of two high-mass stars.  We will see that the causal 
limit on the dimensionless deformability $\Lambda$ is a monotonically decreasing function of $M$ 
and is therefore more stringent for higher mass stars.  On the other hand, the fraction of matter 
above nuclear density is smaller in a low-mass neutron star, and that fact limits the effect of 
a causal EOS above nuclear density.  The net result is that the limit on $\Lambda$ 
set by causality is close to the values of $\Lambda$ associated with candidate neutron-star 
EOSs for matching densities near nuclear density.

\subsection{Effect of Matching Density on Constraints}

To understand the results we present in this section, it is helpful first to 
consider models for which the causal form \eqref{e:causal_eos} extends 
to the surface of the star, where $p=0$. (Here we follow Brecher and Caporaso \cite{Obese} and Lattimer \cite{Lattimer}.)  That is, we consider models based on the EOS 
\be
   p = \epsilon - \epsilon_\mathrm{S},
\label{e:causal_cutoff}\ee
and having finite energy density $\epsilon_S$ at the surface of the star. 
Because the only dimensionful constant is $\epsilon_S$, having 
(in gravitational units) dimension {\it length}$^{-2}$, and the 
mass $M$ and radius $R$ each have dimension {\it length}, 
we have the exact relations 
\be
   M_{\rm max}\propto \epsilon_\mathrm{S}^{-1/2}, \qquad R_{\rm max} \propto \epsilon_\mathrm{S}^{-1/2},  
\ee
where $R_{\rm max}$ is the maximum radius among models with central 
density greater than $\epsilon_{\rm nuc}$ (low-density models have larger radii).
Because the deformability $\lambda$ has dimension ${\it length}^5$, 
we similarly have 
\be
   \lambda_{\rm max}\propto \epsilon_\mathrm{S}^{-5/2}.  
\ee
Using this truncated causal EOS is equivalent 
to taking $\epsilon_\mathrm{S} = \epsilon_{\rm match}$ in the matched causal EOS 
and discarding the envelope of the star below $\epsilon_{\rm match}$.

We emphasize that the truncated EOS (\ref{e:causal_cutoff})
is used only heuristically, to explain the near power-law dependence 
on $\epsilon_{\rm match}$ of the maximum mass, radius, and deformability. (The exact dependence of the maximum mass, radius, and deformability on $\epsilon_{\mathrm{match}}$ is reported below.) 
Because the truncated EOS sets the pressure to zero below $\epsilon_{\rm match}$, 
it underestimates the maximum radius and deformability.   
As noted earlier, to obtain a conservatively large upper limit on maximum deformability, 
we use the matched causal EOS (\ref{e:EOS}), which has a stiff candidate EOS 
for $\epsilon < \epsilon_{\rm match}$.%
\footnote{
There is something paradoxical in using the truncated EOS (\ref{e:causal_cutoff}) 
as an approximation to the EOS that gives the largest possible neutron stars: 
As Lattimer \cite{Lattimer} points out (following Koranda {\it et al.} \cite{KSF}), 
this same EOS gives maximally compact neutron stars, stars with the {\it smallest} 
possible radius for a given mass, among all EOSs consistent with a maximum mass at 
or above a largest observed value, $M_{\rm observed}$. In these and other 
papers \cite{HZ,LP}, \eqcite{e:causal_cutoff} is chosen so that the softest possible EOS (namely $p = 0$) is used up to high density; the stiff causal EOS above that 
density then allows $M_{\rm max} \geq M_{\rm observed}$.  The resolution is this: 
For a {\it fixed maximum mass}, \eqcite{e:causal_cutoff} yields neutron stars with the smallest possible radii.  On the other hand, {\it for a fixed $\epsilon_{\mathrm{match}}$} (i.e. for a given density up to which we assume a known EOS), \eqcite{e:EOS} gives neutron stars with the largest possible radii; and, for low matching and surface densities, the difference between the matched causal EOS (\ref{e:EOS}) and the truncated EOS (\ref{e:causal_cutoff}) becomes negligible. Equivalently,
 $\epsilon_{\mathrm{match}} \rightarrow 0$ 
corresponds to $M_{\mathrm{max}} \rightarrow \infty$, and the difference between
the softest and the stiffest possible EOSs vanishes as $M_{\mathrm max}\rightarrow\infty$. Physically, this happens because a stiff EOS is required to support large masses. 
} 

For $\epsilon_{\rm match} \lesssim \epsilon_{\rm nuc}$, where 
\be
  \epsilon_{\rm nuc} = 2.7\times 10^{14}\  \mathrm{g/cm^3}
\ee
is nuclear saturation density (the central density of large nuclei), 
the contribution of the envelope to mass and radius is small enough that 
the dependence on $\epsilon_{\rm match}$ is very nearly the dependence on 
$\epsilon_\mathrm{S}$ in the truncated star:  
$M_{\rm max}$ and $R_{\rm max}$ are each nearly proportional to 
$\epsilon_{\rm match}^{-1/2}$, and $\lambda_{\rm max}$ is nearly proportional to 
$\epsilon_{\rm match}^{-5/2}$, where 
$M_{\rm max}$ is the maximum neutron-star mass consistent with causality and 
with a low density EOS below $\epsilon_{\rm match}$; and $R_{\rm max}$ and $\lambda_{\rm max}$ are again the corresponding maximum radius 
and deformability among models with central density greater than
$\epsilon_{\rm nuc}$. This behavior can be seen in Figs.~\fignumcite{fig:logMmaxandlogRmax}
and \fignumcite{fig:loglambdamax}, where linear least-squares fits to the leftmost 10 data points in each plot satisfy 
\begin{subequations}\begin{align} 
M_{\rm max} &= (4.1 \ M_{\odot}) (\epsilon_{\rm match}/\epsilon_{\rm nuc})^{-.4999},\\
R_{\rm max} &= (17  \ \mathrm{km}) (\epsilon_{\rm match}/\epsilon_{\rm nuc})^{-.4990}, \\
\lambda_{\rm max} &= (1.3 \times 10^{37} \mathrm{g\ cm^2\ s^2})(\epsilon_{\rm match}/\epsilon_{\rm nuc})^{-2.4996} .
\label{e:lambda_max}\end{align}
\label{e:MRlambda_max}\end{subequations}
\begin{figure*}
	\subfloat[\label{fig:logMmaxandlogRmax}]{\includegraphics[height=.33\textheight]{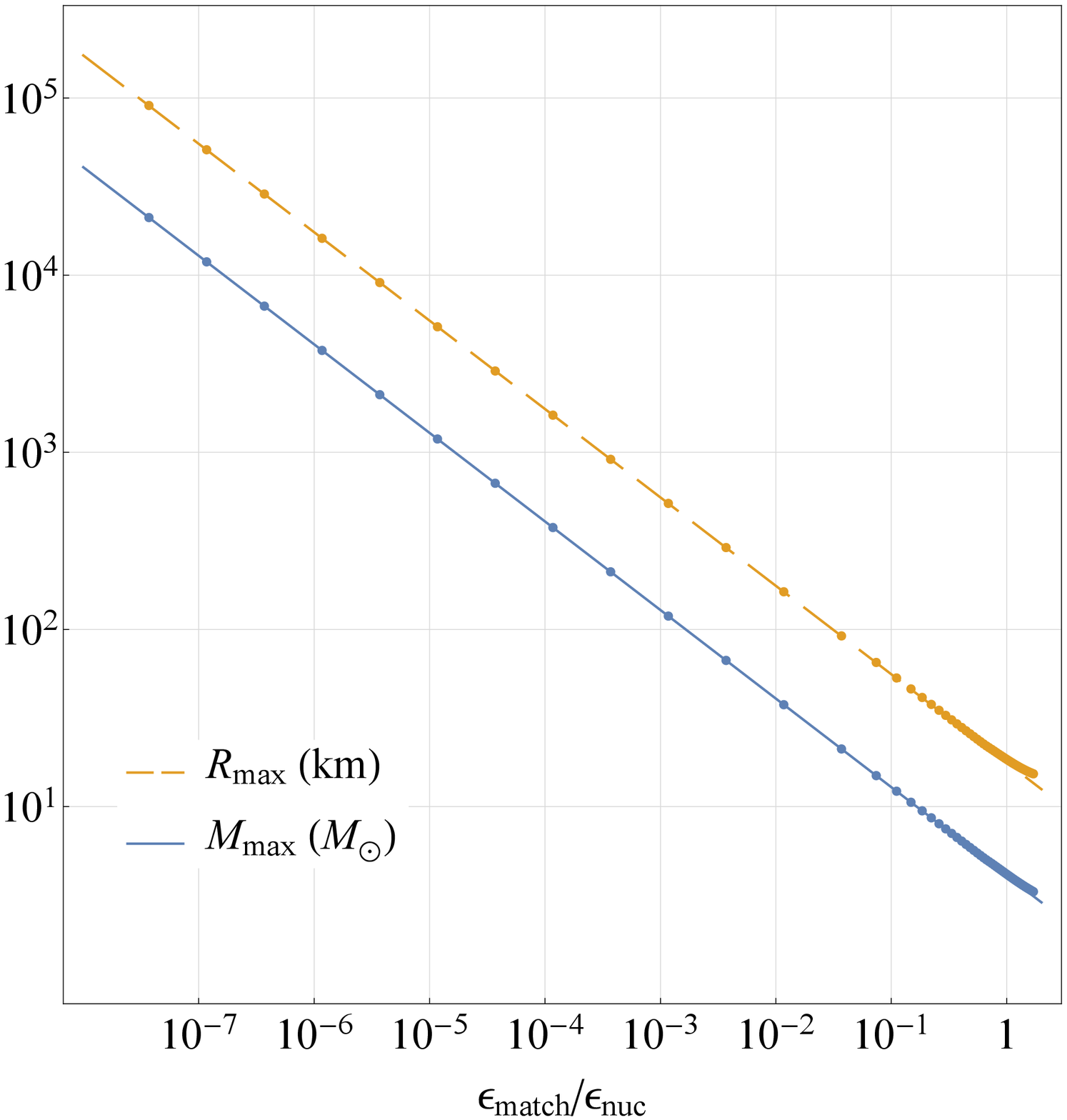}}		
	\hspace{.06\textwidth}
	\subfloat[\label{fig:loglambdamax}]{\includegraphics[height=.33\textheight]{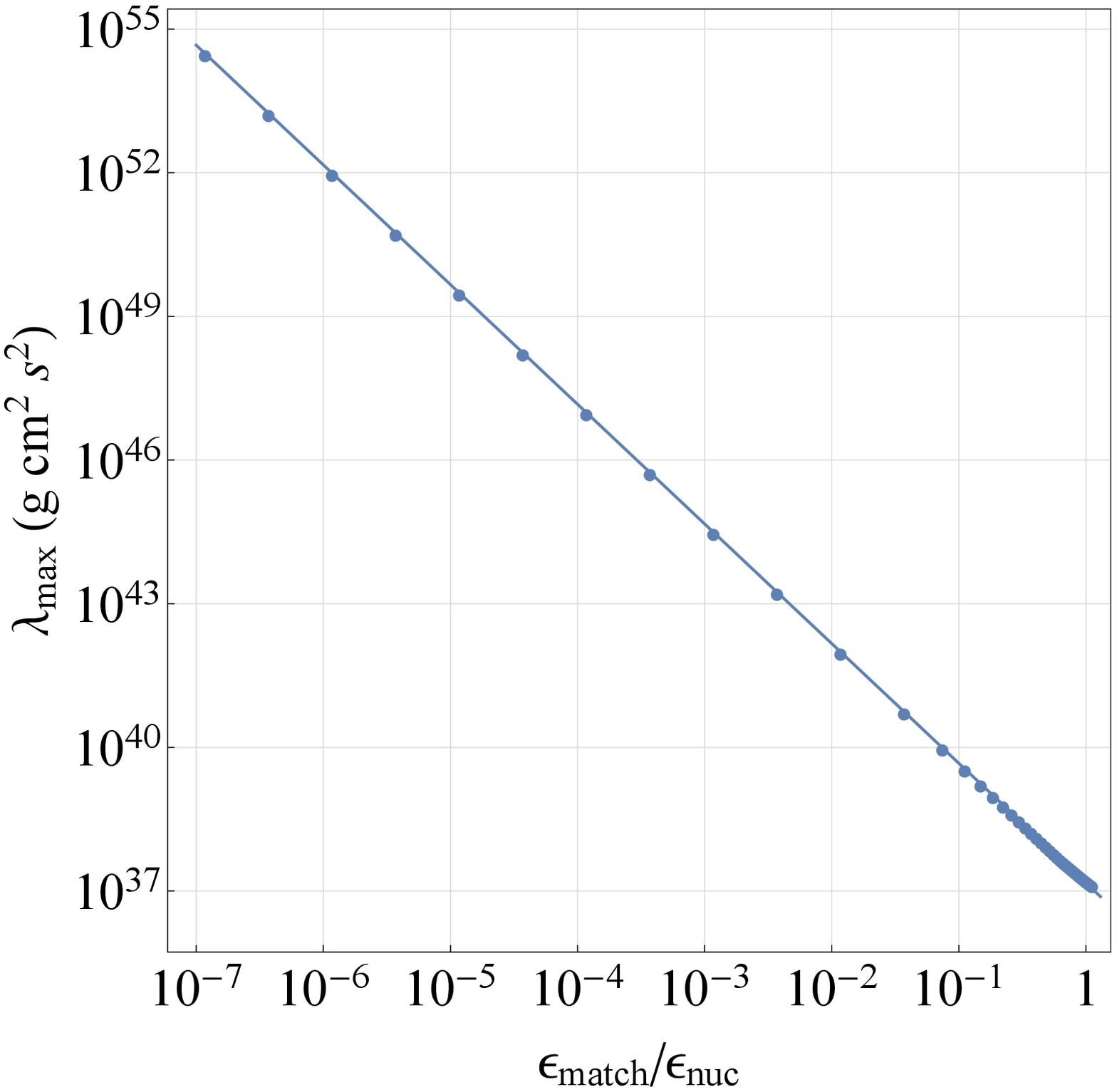}}	
	\caption{The (a) maximum radius, mass, and (b) tidal deformability are plotted against the matching density. The behavior of all three quantities follow a power law except at high $\epsilon_{\rm match}$, with the best-fit lines given by \eqcite{e:MRlambda_max}.} 
\end{figure*}
The rightmost data points in each plot diverge from the line because, at higher matching densities, a larger envelope obeys the low-density (MS1) EOS.

\begin{figure}
	\includegraphics[width = \columnwidth]{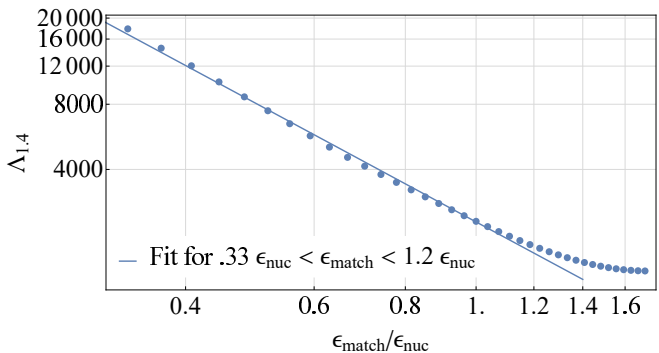}
	\caption{The dependence of the dimensionless tidal deformability $\Lambda_{1.4}$ of $1.4 M_{\odot}$ stars on matching density is shown on a log-log plot. The behavior approximates 
		a power law for $\epsilon \lesssim \epsilon_{\rm nuc}$, with the best fit given by \eqcite{logLambda1-4}. \label{fig:Lambda1-4vsepsilon}}
\end{figure}

Of greater astrophysical relevance than the upper limit on $\lambda$, however,
is the constraint on the dimensionless 
tidal deformability, $\Lambda = \lambda/M_{\mathrm{NS}}^5 = \frac23 k_2 R^5/M^5$, 
that governs the waveform of a binary inspiral.  
As we will see below, because of the factor $M_{\mathrm{NS}}^{-5}$,  
$\Lambda$ is monotonically decreasing with increasing mass for central density above 
$\epsilon_{\rm match}$.  The physically interesting constraint on $\Lambda$ is then a 
constraint at known mass:  Inspiral waveforms detected with a high enough 
signal-to-noise ratio to measure their tidal departure from point-particle inspiral 
will also have the most accurately measured neutron-star masses.  The dependence of 
$\Lambda$ on $\epsilon_{\rm match}$ for fixed mass cannot be found from the 
previous dimensional analysis, but it is easy to see that $\Lambda(M,\epsilon_{\rm match})$ is a monotonically decreasing function of $\epsilon_{\rm match}$:  
As $\epsilon_{\rm match}$ increases and less of the 
star is governed by the stiffer causal EOS, the star becomes more compact: 
$R$ decreases at fixed $M$.  In addition, as the density profile becomes more 
centrally condensed, the tidal Love number $k_2$ decreases, because, for a given radius, the external tidal force has less effect on a more centrally condensed star.  
Decreasing $R$ and $k_2$ gives a sharp decrease in $\Lambda$, as shown in \figcite{fig:Lambda1-4vsepsilon} for a $1.4 M_\odot$ star.  
For $.33\epsilon_{\rm nuc} < \epsilon_{\rm match} < 1.2\epsilon_{\rm nuc} $ we find a near power-law dependence, 
\be
\Lambda_{1.4} = 2400 (\epsilon_{\rm match}/\epsilon_{\rm nuc})^{-1.8}.
\label{logLambda1-4}\ee

\subsection{Comparison between Constraint and Results from Candidate EOSs}

\begin{figure*}[!ht]
	\subfloat[\label{MR}]{\includegraphics[width=.45\textwidth]{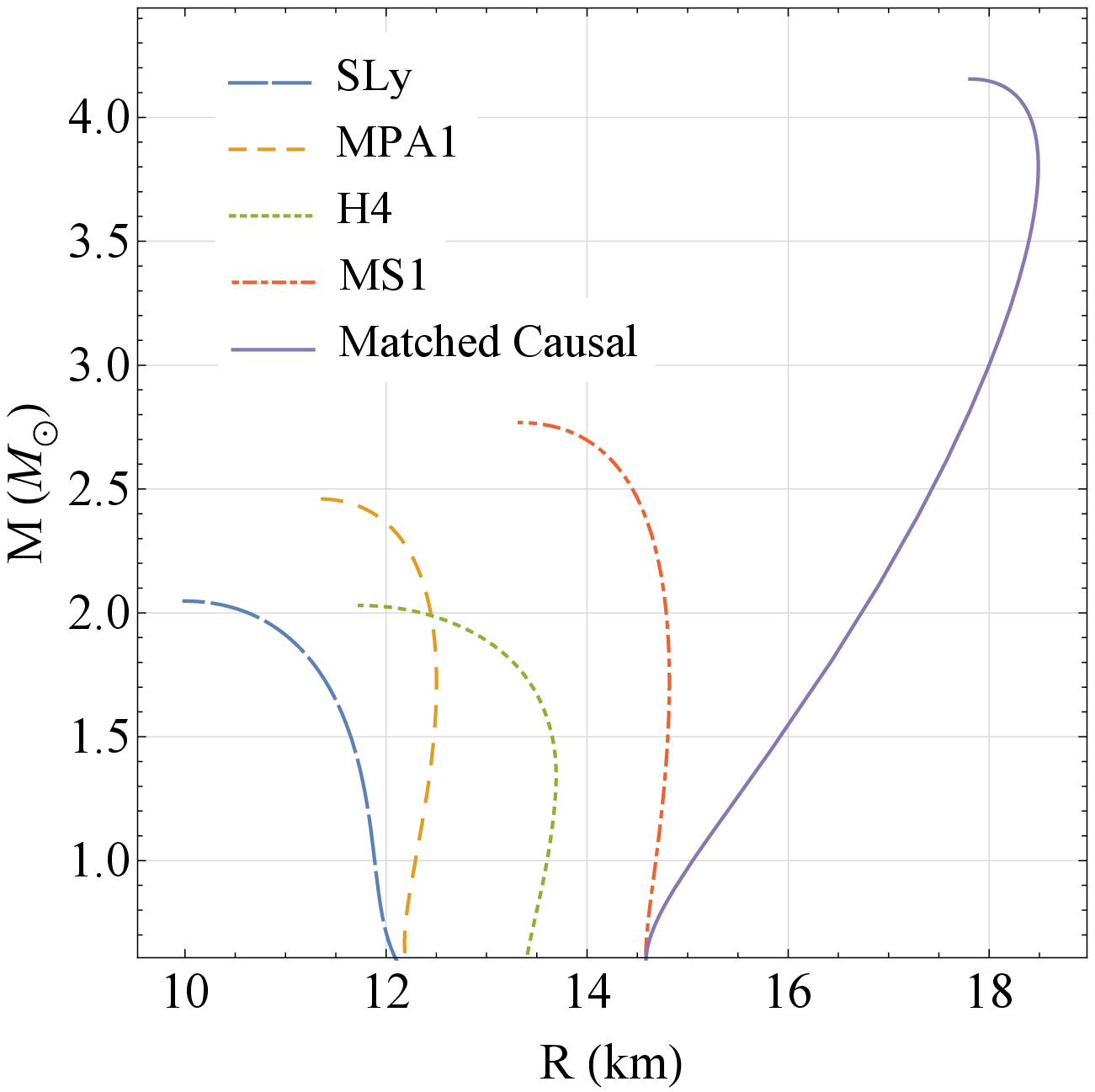}}
	\hspace{.06\textwidth}
	\subfloat[\label{lambdavsM}]{\includegraphics[width=.45\textwidth]{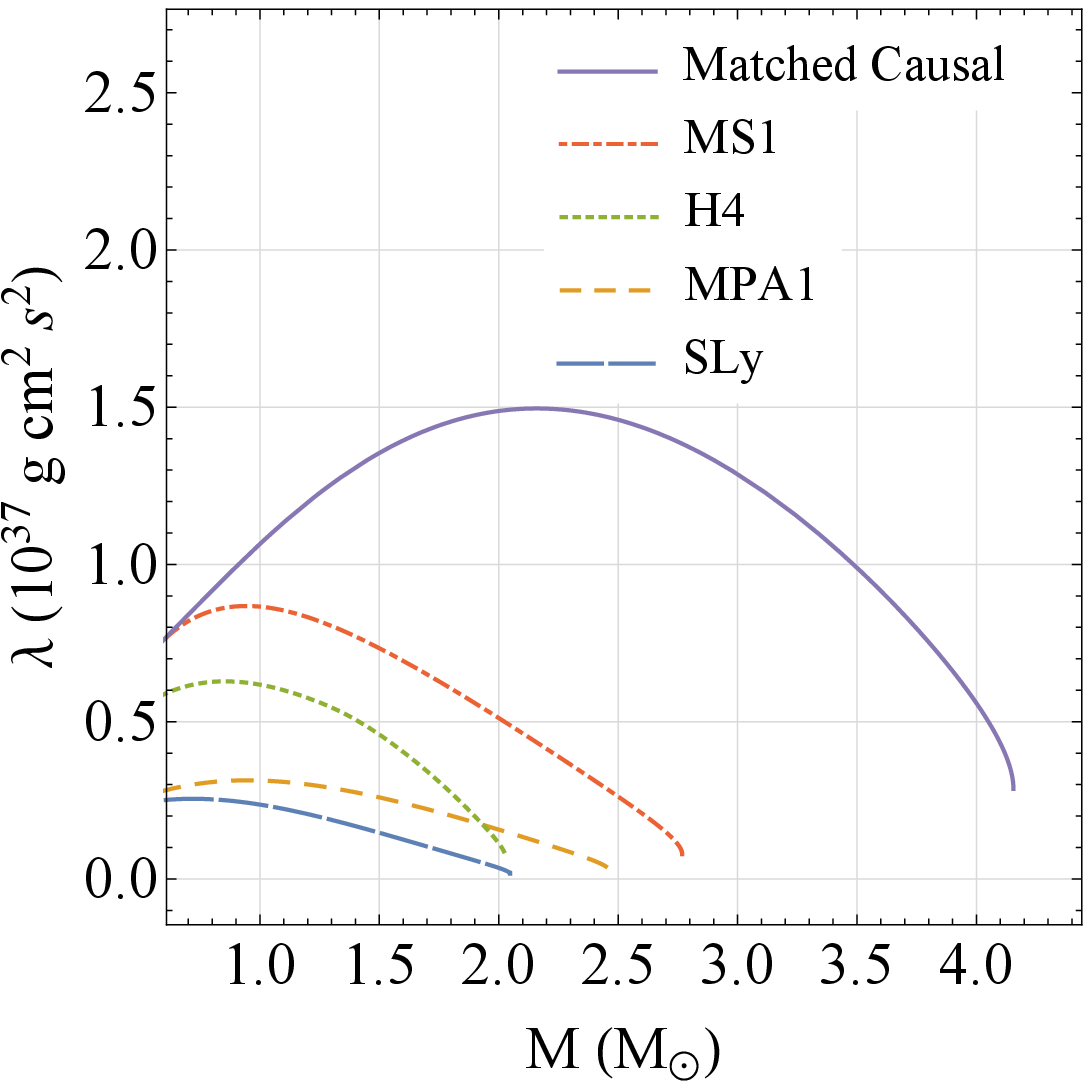}}
	\caption{(a) The mass-radius relation for the matched causal EOS with $\epsilon_{\rm match}=\epsilon_{\rm nuc}$ and for candidate neutron-star equations 
		of state that display the range of uncertainty in stiffness.\\
		(b) Tidal deformability versus mass for stars based on the same EOSs. The top solid curve, displaying the tidal deformability of stars based on the matched causal EOS, is an upper limit set by causality on tidal deformability. Stars based on softer EOSs 
		have smaller tidal deformabilities.}
	\label{fig:MR_lambdavsM}
\end{figure*}

We begin by displaying the limit set by causality on the dimensionful tidal
deformability $\lambda$ as a function of mass, with $\epsilon_{\rm match}$ taken to be 
$\epsilon_{\rm nuc}$.   There is remaining uncertainty in the equation of 
state at $\epsilon_{\rm nuc}$, and we obtain a conservative upper limit by matching to 
the MS1 EOS \cite{MS1}, which is particularly stiff for $\epsilon \lesssim \epsilon_{\mathrm{nuc}}$.

The mass-radius relation for the family of neutron stars obeying the matched causal EOS is  indicated by ``Matched Causal'' in Fig.~\fignumcite{MR}.  As we saw in  
Eqs.~(\ref{e:MRlambda_max}), matching to MS1 below  $\epsilon_{\rm nuc}$ is a weak 
constraint, giving $M_{\rm max}=4.1 M_{\odot}$ and $R_{\rm max}>18\ \mathrm{km}$, both significantly larger than their values for any of the candidate EOSs shown. These candidate EOSs include SLy \cite{SLy}, which is one of the softest EOSs that allow for $2 M_{\odot}$ neutron stars, MPA1 \cite{MPA1}, which is slightly stiffer, H4 \cite{H4}, which is stiff at low densities and soft at high densities, and MS1 \cite{MS1}, which is particularly stiff at all densities. The maximum masses allowed by these EOSs are all between $2$ and $2.8\ M_{\odot}$, and the radii are all between $10$ and $15\ \mathrm{km}$.

\begin{figure*}
	\subfloat{\includegraphics[width = .75\textwidth]{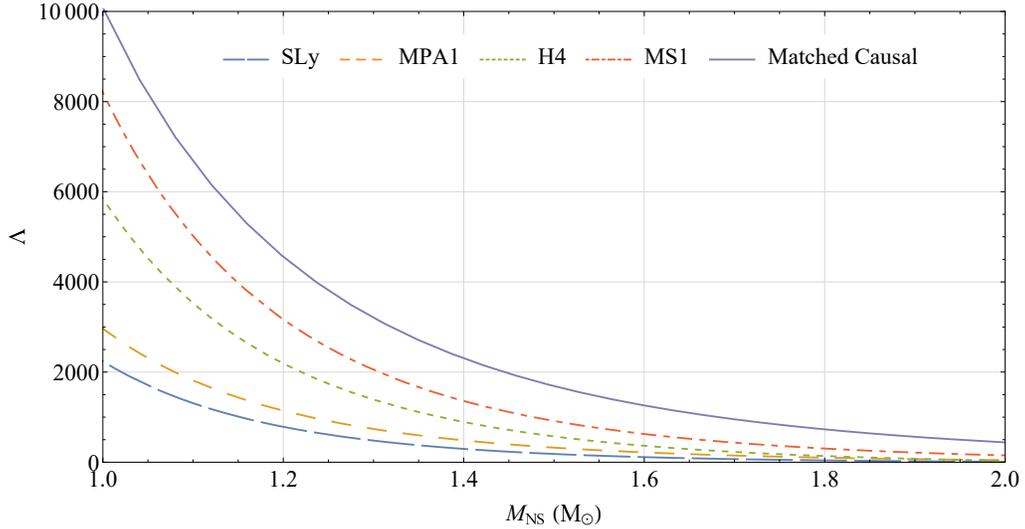}}
	\caption{The dimensionless tidal deformability $\Lambda$ is plotted against mass for several EOSs. For any given mass, the Matched Causal EOS places an upper limit on the value of $\Lambda$. \label{fig:Lambda}}
\end{figure*}

In Figure \fignumcite{lambdavsM}, the top curve displays an  
upper limit on $\lambda$ as a function of neutron-star mass obtained from the 
matched causal EOS. The comparison $\lambda(M)$ curves for the same candidate EOSs of Fig.~\fignumcite{MR} show the decreasing deformability associated with 
stars of decreasing stiffness and radius.  Note, however, that the maximum value of 
$\lambda$ for each EOS occurs at a smaller mass than that of the model with 
maximum radius. 
This is due to the increase in central condensation as the mass increases, resulting 
in an decrease in $k_2$. The maximum of the $\lambda(M)$ curve for the matched causal EOS gives the mass-independent upper limit $\lambda < 1.5 \times 10^{37}\ \mathrm{g\ cm^2\ s^2}$, for 
$\epsilon_{\rm match}=\epsilon_{\rm nuc}$, with dependence on $\epsilon_{\rm match}$ 
given by Eq.~(\ref{e:lambda_max}) for smaller matching density.    


The corresponding upper limit $\Lambda_{\rm max}(M)$ on the dimensionless deformability  is given by the top curve in \figcite{fig:Lambda}, for $\epsilon_{\rm match}=\epsilon_{\rm nuc}$.
(The dependence on $\epsilon_{\rm match}$ was shown in 
Fig.~\ref{fig:Lambda1-4vsepsilon} for a representative $1.4 M_\odot$ star.)  Since $\Lambda \propto C^{-5}$, $\Lambda$ is large for small masses and relatively small for larger masses. As a result, it is not meaningful to speak of a 
mass-independent maximum of $\Lambda$,
but it is meaningful to compare $\Lambda$-values at constant mass. The most striking feature of \figcite{fig:Lambda} is how close the curve $\Lambda_{\rm max}(M)$ 
is to the range of $\Lambda$ allowed by current candidate EOSs. 
This stringent constraint on $\Lambda$ is in sharp contrast to the larger departures 
of the curves giving $R_{\rm max}(M)$ and $\lambda_{\rm max}(M)$ in \figcite{fig:MR_lambdavsM} 
from the corresponding curves for candidate EOSs.  
For $1.4\ M_{\odot}$ stars, for example, it places the constraint that $\Lambda \leq 2300$. For comparison, $1.4\ M_{\odot}$ stars resulting from the SLy, MPA1, H4, and MS1 EOSs have $\Lambda$-values of 300, 490, 900, and 1400, respectively.

	\begin{figure*}
	\centering
	\includegraphics[width=.75\textwidth]{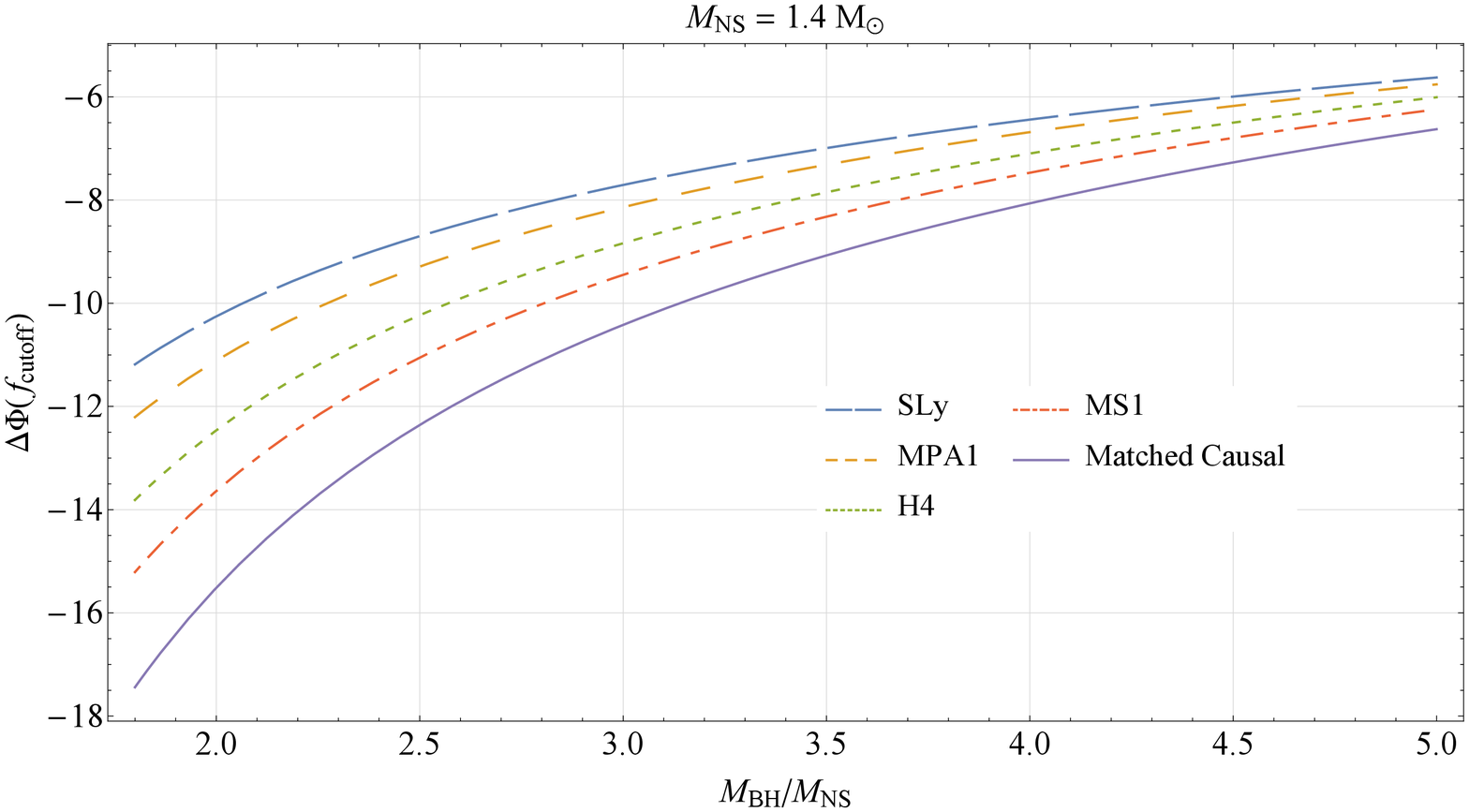}
	\caption{The estimated total gravitational wave phase shift $\Delta\Phi(f_{\rm cutoff})$ corresponding to a BHNS binary with $M_{\mathrm{NS}} = 1.4\ M_{\odot}$ and $\chi_{\mathrm{BH}} = 0$ is plotted against the mass ratio for several EOSs. For a given mass ratio, $|\Delta\Phi(f_{\rm cutoff})|$ is larger for stiffer EOSs, and the Matched Causal EOS provides a constraint on it. In general, $|\Delta\Phi(f_{\rm cutoff})|$ decreases with the mass ratio. \label{fig:DeltaPhiMaxvsQDiffEOS}}
	\end{figure*}

	One might naively expect $|\Delta\Phi(f_{\rm cutoff})|$ to increase monotonically with $\Lambda$ and therefore to decrease monotonically with the mass $M_{\mathrm{NS}}$ of the neutron star (note that, although $\Delta\Phi$ is positive when evaluated at a given time, it is negative when evaluated at a given frequency). This is not the case, because while $|\Delta\Phi|$ increases with $\Lambda$ (and decreases with $M_{\mathrm{NS}}$) when evaluated at a fixed frequency, $f_{\rm cutoff}$ decreases monotonically with $\Lambda$ (and increases monotonically with $M_{\mathrm{NS}}$). That is, stars with high dimensionless tidal deformability are tidally disrupted at a larger distance from the black hole, corresponding to a smaller orbital (and gravitational wave) frequency. A neutron star with high tidal deformability therefore has fewer cylces during which to accumulate phase relative to a point-particle. As a result, the effect of $M_{\mathrm{NS}}$ on $|\Delta\Phi(f_{\rm cutoff})|$ is complicated, and depends on EOS and the parameters of the binary.

	\begin{figure*}
		\centering
		\includegraphics[width=.75\textwidth]{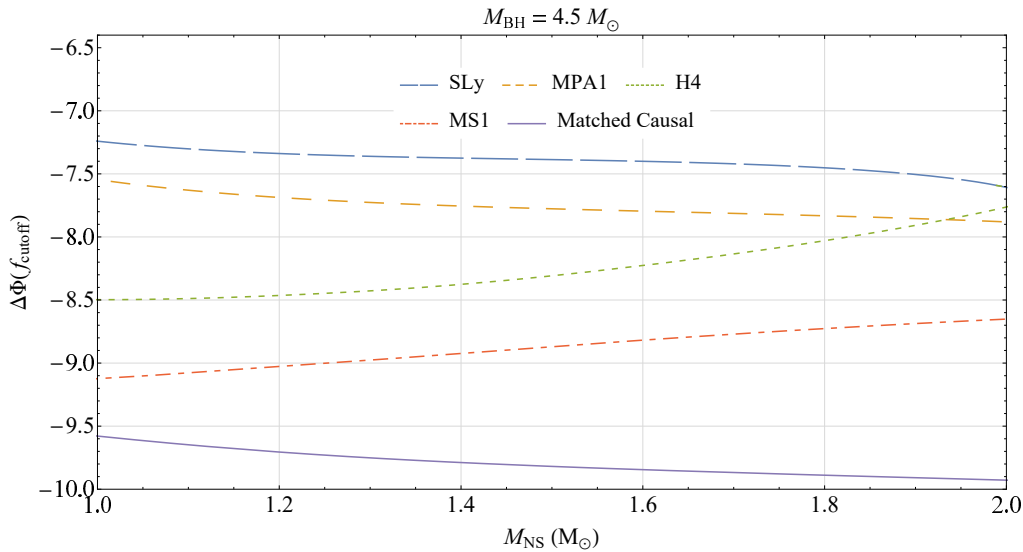}
		\caption{The estimated 
total gravitational wave phase shift $\Delta\Phi(f_{\rm cutoff})$ corresponding to a BHNS binary with $M_{BH} = 4.5\ M_{\odot}$  and $\chi_{\mathrm{BH}} = 0$ is plotted against neutron star mass for several EOSs. For a given mass, $|\Delta\Phi(f_{\rm cutoff})|$ is larger for stiffer EOSs, and the Matched Causal EOS provides a constraint on it. The dependence of $|\Delta\Phi(f_{\rm cutoff})|$ on neutron star mass is complicated and changes with the EOS used. \label{fig:DeltaPhiMaxvsMdiffEOS}}
	\end{figure*}

Nevertheless, stiffer EOSs result in larger values of $|\Delta\Phi|$ for given neutron star masses or mass ratios. As can be seen in \figcite{fig:DeltaPhiMaxvsQDiffEOS}, $|\Delta\Phi(f_{\rm cutoff})|$ decreases with mass ratio for all EOSs. On the other hand, $|\Delta\Phi(f_{\rm cutoff})|$ has complicated behavior with respect to neutron star mass for all EOSs when the spin of the companion black hole is zero (\figcite{fig:DeltaPhiMaxvsMdiffEOS}). In addition, one can see in \figcite{fig:DeltaPhiMaxvsQDiffEOS} and \figcite{fig:DeltaPhiMaxvsMdiffEOS} that $|\Delta\Phi(f_{\rm cutoff})|$ increases with the stiffness of the EOS, and is largest for our EOS, but only by a few radians at most. 
Here, based on our estimate of $\Delta\Phi$, the constraint set by causality 
is remarkably strong, stronger than the already stringent constraint on $\Lambda$: 
$\Delta\Phi_{\rm max}(M)$ differs from its value for the stiffest candidate equation 
of state by less than 14\%. 
The strength of the causal constraint is due to (a) the fact that $\Lambda$ 
is largest at small mass, where the causal EOS governs the smallest 
fraction of the star, and (b) a smaller cutoff frequency for the stiffest EOSs that reduces the time over which the phase can accumulate.


	As shown in \figcite{fig:DeltaPhiMaxvsM}, 
for a given black hole mass $M_{\mathrm{BH}}$ and zero black hole spin $\chi_{\mathrm{BH}}$, 
$|\Delta\Phi(f_{\rm cutoff})|$  
increases with $M_{\mathrm{NS}}$ for the Matched Causal EOS. In addition, for a given $M_{\mathrm{NS}}$, 
$|\Delta\Phi(f_{\rm cutoff})|$ decreases with $M_{\mathrm{BH}}$. Changing $\chi_{\mathrm{BH}}$ can change 
the qualitative behavior of $|\Delta\Phi(f_{\rm cutoff})|$, as can be seen in 
\figcite{fig:DeltaPhiMaxvsMdiffChi}. In particular, a corotating companion black hole tends to make 
$|\Delta\Phi(f_{\rm cutoff})|$ increase with mass, while antirotating companions tend to make 
$|\Delta\Phi(f_{\rm cutoff})|$ decrease with mass. For a given $M_{\mathrm{NS}}$, higher (corotating) 
spins result in smaller $|\Delta\Phi(f_{\rm cutoff})|$, but the effect decreases with increasing 
$M_{\mathrm{NS}}$.

	\begin{figure*}
		\centering
		\includegraphics[width=.75\textwidth]{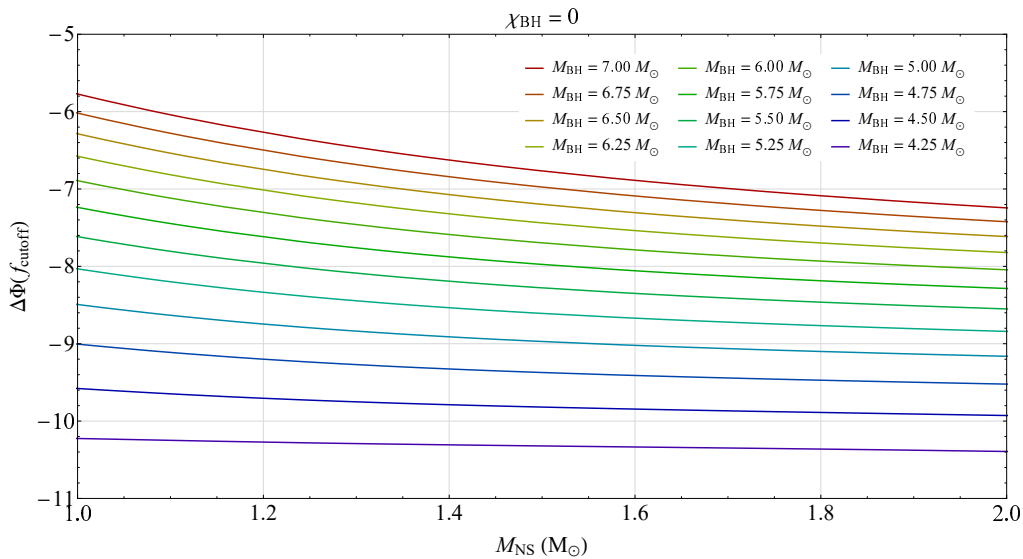}
		\caption{The estimated constraint on $\Delta\Phi(f_{\rm cutoff})$ is plotted against the mass of a neutron star for several different black hole masses and a black hole spin of 0. We expect that the absolute value of $\Delta\Phi$ would be lower for any real BHNS binary. The constraint on $|\Delta\Phi|$ decreases with both neutron star mass and black hole mass. \label{fig:DeltaPhiMaxvsM}}
	
	\end{figure*}

	\begin{figure*}
		\centering
		\includegraphics[width=.75\textwidth]{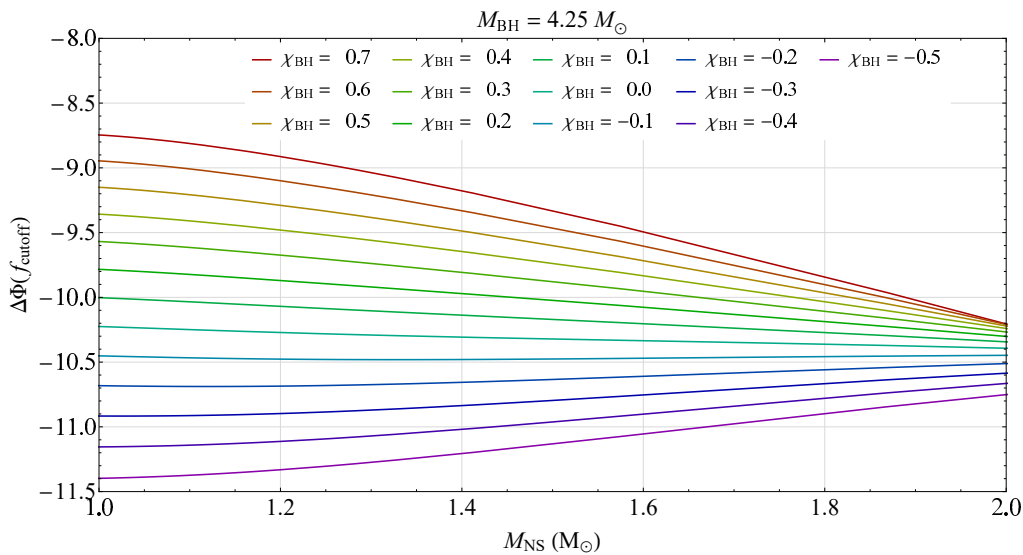}
		\caption{The estimated constraint on $\Delta\Phi(f_{\rm cutoff})$ is plotted against the mass of a neutron star for several different black hole spins and a black hole mass of $4 M_\odot$. Different black hole spins can change how $\Delta\Phi$ qualitatively changes with neutron star mass, and $\Delta\Phi$ depends more strongly on $\chi_{BH}$ for smaller neutron star masses than for larger neutron star masses. \label{fig:DeltaPhiMaxvsMdiffChi}}
	
	\end{figure*}
	
	Figure \fignumcite{fig:DeltaPhiMaxvsQdiffMns} shows how $|\Delta\Phi(f_{\rm cutoff})|$ varies with mass ratio for several neutron star masses and 0 black hole spin. For a given $M_{\mathrm{NS}}$, $|\Delta\Phi(f_{\rm cutoff})|$ decreases with increasing mass ratio. For a given mass ratio, $|\Delta\Phi(f_{\rm cutoff})|$ decreases with neutron star mass. The effect decreases in magnitude as the mass ratio increases.

 Finally, \figcite{fig:DeltaPhiMaxvsQdiffChi} shows $|\Delta\Phi(f_{\rm cutoff})|$ varying with mass ratio 
for several black hole spins and $M_{\mathrm{NS}} = 1.4\ M_{\odot}$. $|\Delta\Phi(f_{\rm cutoff})|$ 
decreases with mass ratio regardless of the value of $\chi_{\mathrm{BH}}$, but for a given mass ratio, 
$|\Delta\Phi(f_{\rm cutoff})|$ decreases with $\chi_{\mathrm{BH}}$; it is smallest for corotating black 
holes, and largest for antirotating black holes.

	\begin{figure*}
		\centering
		\includegraphics[width=.75\textwidth]{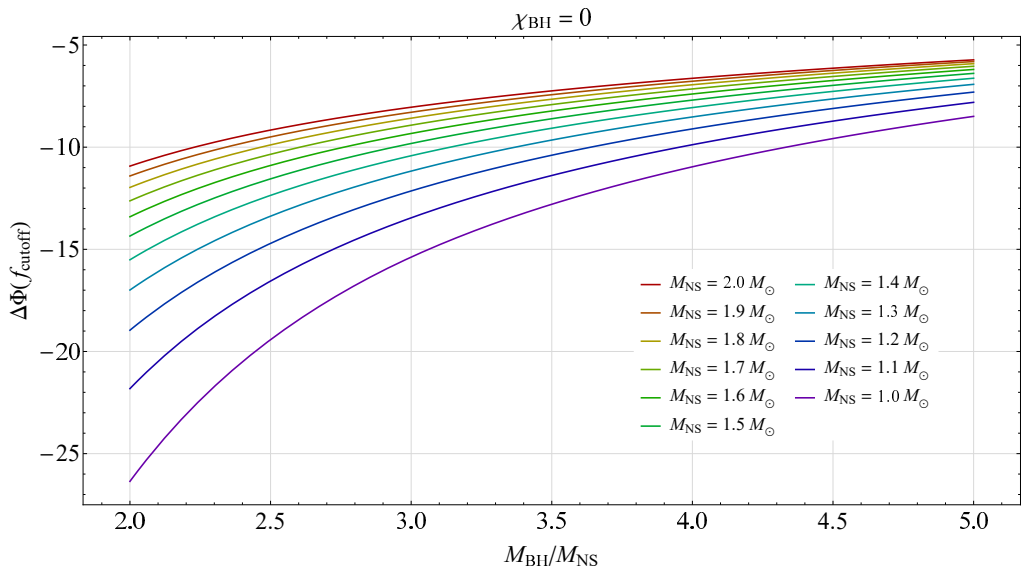}
		\caption{The estimated constraint on $|\Delta\Phi(f_{\rm cutoff})|$ for BHNS binaries with $\chi_{BH} = 0$ is plotted against the mass ratio for several neutron star masses. $|\Delta\Phi(f_{\rm cutoff})|$ decreases both with mass ratio and with neutron star mass. \label{fig:DeltaPhiMaxvsQdiffMns}}	
	\end{figure*}

	\begin{figure*}
		\centering
		\includegraphics[width=.75\textwidth]{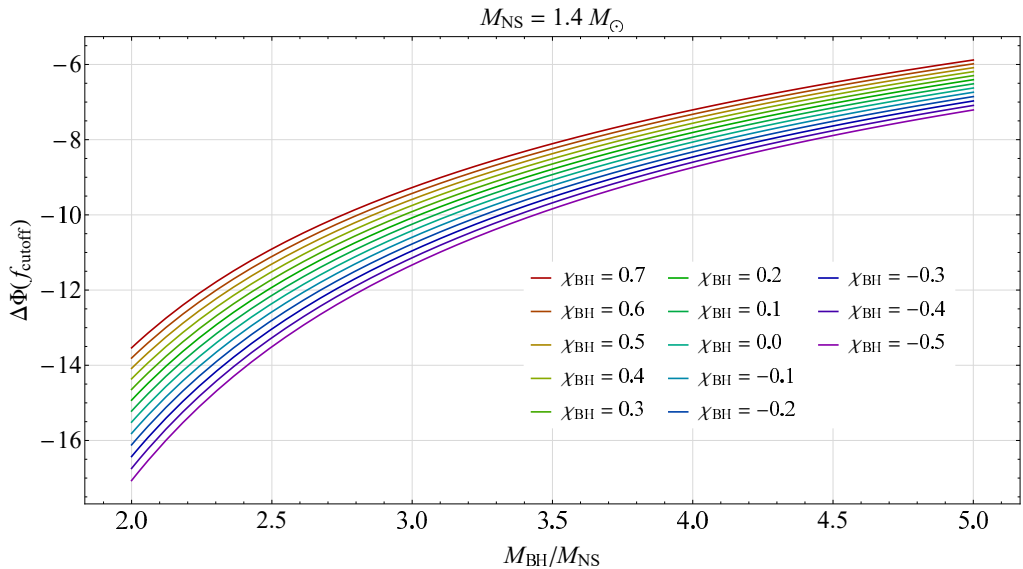}
		\caption{The estimated constraint on $|\Delta\Phi(f_{\rm cutoff})|$ for BHNS binaries with $M_{NS} = 1.4\ M_{\odot}$ is plotted against the mass ratio for several black hole spins. The value of $|\Delta\Phi(f_{\rm cutoff})|$ decreases with mass ratio and with spin. \label{fig:DeltaPhiMaxvsQdiffChi}}	
	\end{figure*}

	\section{Conclusion}
By using a stiffest causal EOS consistent with causality at high density, 
matched to the MS1 \cite{MS1} EOS below a density $\epsilon_{\rm match}$, 
we have set upper limits on the quadrupole tidal deformability 
$\lambda$ and on the dimensionless tidal deformability $\Lambda$ 
as a function of neutron star mass.  
The limit on $\Lambda$, given by Eq.~(\ref{logLambda1-4}) for a 1.4 $M_\odot$ neutron star, 
is conservative, because we have matched to an EOS (MS1) that is 
stiff below nuclear density: With this low-density EOS and a match at 
$\epsilon_{\rm nuc}$, the corresponding upper mass limit is $4.1 M_\odot$. 
Using the constraint on dimensionless tidal deformability and the Lackey {\it et al.}
analytic fit to numerical data \cite{Lackey}, we then estimated the induced phase shift of a BHNS inspiral and merger waveform. 

The implied upper limit on the accumulated phase shift 
$|\Delta\Phi|$ at merger depends on the parameters of the binary, but it is surprisingly 
close to the range of phase shifts seen in candidate EOSs. Assuming one 
can neglect resonant interactions of the tidal field with neutron-star modes, we think 
this conclusion is secure.  We emphasize, however, that our upper limits on $|\Delta\Phi|$ rely on an analytic expression based on full numerical simulations for models with a set of 
EOSs significantly less stiff than the matched causal EOS.  
Work {\it has} begun on numerical simulations to obtain an upper limit on the departure of 
double neutron star inspiral waveforms from the point-particle (or spinless BBH) case.

	\begin{acknowledgments}
	We would like to thank Benjamin Lackey and Lee Lindblom for sharing data with us, Lindblom for a useful discussion regarding multiple-parameter EOSs, and James Lattimer for 
helpful comments on the paper.  
	This work was supported in part by NSF Grant Nos. PHY-1307429 and PHY-1607585.
	\end{acknowledgments}

\bibstyle{prd}
\normalem
\bibliography{TidalDeformabilityBib}

\appendix
\section{Comments on causality}
\label{s:causality}

With the assumption that the equilibrium equation of state of the neutron star 
and its perturbations are governed by the same one-parameter equation of state, 
causality implies $dp/d\epsilon < 1$.  That is, as mentioned in the text, 
the time-evolution of a barotropic fluid is described by a hyperbolic system whose 
characteristics lie within the light cone precisely when $dp/d\epsilon<1$
\cite{lichnerowicz}.  The frequencies of stellar perturbations, 
however, are too high for the temperature of a fluid element and the relative 
density $Y_i$ of each species of particle to reach their values for the 
background fluid at the same pressure: Heat flow and nuclear reactions 
are incomplete.  

Because of this, one cannot precisely identify the maximum 
speed of signal propagation in the fluid with the equilibrium value
\[
   \sqrt{\left.\frac{dp}{d\epsilon}\right|_{\rm equilibrium}} 
		:= \sqrt{\frac{dp/dr}{d\epsilon/dr}}.
\] 
If short wavelength, high frequency perturbations are too rapid for heat 
flow and for nuclear reactions to proceed, their speed of propagation is  
\be
v_{\rm sound} = \sqrt{(\partial p/\partial\epsilon)|_{s,Y_i}}.
\label{e:vsound} 
\ee
One therefore expects causality to imply 
\be
 \left.\frac{\partial p}{\partial\epsilon}\right|_{s,Y_i} < 1.     
\label{e:vssY}\ee
This is known to be true for a relativistic fluid with a two-parameter EOS 
of the form $p = p(\epsilon,s)$:  Its dynamical evolution then involves heat 
flow and is governed by the equations of a dissipative relativistic fluid.
Causal theories of this kind were first introduced by Israel and Stewart \cite{israel76,IS79} and by Liu {\it et al.} \cite{LMR86}.  The general class of such theories was 
analyzed by Geroch and Lindblom \cite{GL91}, who pointed out that, for dissipative fluids 
obeying $p = p(\epsilon,s)$, causality implies the inequality~(\ref{e:vssY}), 
\be
\left.\frac{\partial p}{\partial\epsilon}\right|_{s} < 1.     
\label{e:vss}\ee
Now a star is unstable to convection if 
\be
  \left.\frac{dp}{d\epsilon}\right|_{\rm equilibrium} 
		 > \left.\frac{\partial p}{\partial\epsilon}\right|_{s,Y_i}.
\ee
Thus, {\it for a locally stable spherical star (a self-gravitating equilibrium configuration 
of a relativistic dissipative fluid) based on a two-parameter 
EOS $p = p(\epsilon,s)$, causality implies} 
\be
  \sqrt{\left.\frac{dp}{d\epsilon}\right|_{\rm equilibrium}} < 1. 
\label{e:ineqequil}\ee
Thus, at least for two-parameter dissipative fluids, one can rule out the 
possibility that dispersion in a dissipative fluid could lead to a group velocity 
smaller than the phase velocity (see, for example Bludman and Ruderman \cite{BR70}) 
$v_{\rm sound}$ and thereby allow $v_{\rm sound} >1$ without superluminal 
signal propagation.    

For a dissipative fluid obeying a multi-parameter EOS of the form $p = p(\epsilon,s,Y_i)$,
we are not aware of a general proof that causality implies the inequality 
(\ref{e:vssY}).  One has only the weaker statement, 
{\it for a locally stable spherical star based on an EOS 
equation of state $p = p(\epsilon,s)$, $v_{\rm sound} < 1$ implies 
the equilibrium inequality (\ref{e:ineqequil}).}  
There is one additional caveat: The core of a neutron star is likely to 
be a superfluid, and taking that into account could lead to small corrections 
in the speed of sound.  

Finally, we note that for candidate EOSs, although the inequality $v_{\rm sound} < 1$ 
is stronger than the the equilibrium inequality (\ref{e:ineqequil}) used to place upper 
limits on mass, radius and, in the present paper, on deformability, the difference is 
small.  The fractional difference 
\be
   \frac{\sqrt{dp/d\epsilon|_{\rm equilibrium}} - \sqrt{(\partial p)/(\partial\epsilon)|_{s,Y_i}}}
	{\sqrt{(\partial p)/(\partial\epsilon)|_{s,Y_i}}} 
\ee
is primarily due to composition (to the constant values of $Y_i$), and it  
is less than 5\%. (It is approximately half the fractional difference between the
adiabatic index $\gamma=\Gamma_1$ and the index $\Gamma$ governing the equilibrium configuration; the difference determines the Brunt-Va\"is\"al\"a frequency, a 
characteristic frequency of $g$-modes, and an estimate can be found, for example, in 
Ref.~\cite{GR92}.)

\end{document}